\documentclass[12pt,preprint]{aastex}
\RequirePackage{color}
\usepackage{psfig}
\usepackage{epsfig}
\begin{document}
\newcommand\tbd[1]{\textbf{\color{red}(#1)}}
\def\ale{\mathrel{\hbox{\rlap{\hbox{\lower4pt\hbox{$\sim$}}}\hbox{$<$}}}}
\def\age{\mathrel{\hbox{\rlap{\hbox{\lower4pt\hbox{$\sim$}}}\hbox{$>$}}}}
\def\Lya{{Ly$\alpha$}\/}

\title{Cosmological Implications of the Very High Redshift GRB~050904}

\author{
P.~A.~Price,$\!$\altaffilmark{1}
L.~L.~Cowie,$\!$\altaffilmark{1}
T.~Minezaki,$\!$\altaffilmark{2}
B.~P.~Schmidt,$\!$\altaffilmark{3}
A.~Songaila,$\!$\altaffilmark{1}
Y.~Yoshii$\!$\altaffilmark{2}
}

\altaffiltext{1}{Institute for Astronomy, University of Hawaii, 2680
Woodlawn Drive, Honolulu, HI 96822. {\tt price,cowie,acowie@ifa.hawaii.edu}}

\altaffiltext{2}{Institute of Astronomy, University of Tokyo, Mitaka,
Tokyo 181-0015, Japan. {\tt minezaki,yoshii@mtk.ioa.s.u-tokyo.ac.jp}}

\altaffiltext{3}{Research School of Astronomy and Astrophysics,
Australian National University, via Cotter Road, Weston, ACT, 2611,
Australia. {\tt brian@mso.anu.edu.au}}

\slugcomment{Accepted for publication in The Astrophysical Journal} 

\begin{abstract}
We report near simultaneous multi-color ($RIYJHK$) observations made
with the MAGNUM 2m telescope of the gamma ray burst GRB~050904
detected by the SWIFT satellite. The spectral energy distribution
shows a very large break between the $I$ and $J$ bands.  Using
intergalactic transmissions measured from high redshift quasars we
show that the observations place a 95\% confidence lower limit of
$z=6.18$ on the object, consistent with a later measured spectroscopic
redshift of 6.29 obtained by \citet{2005GCN..3937....1K} with the
Subaru telescope.  We show that the break strength in the $R$ and $I$
bands is consistent with that measured in the quasars. Finally we
consider the implications for the star formation history at high
redshift.
\end{abstract}


\section{Introduction}
\label{secintro}

Identifying the epoch of reionisation remains a prime objective in
modern cosmology.  Quasars at the highest redshifts, $z>6.1$ of which
five are known
\citep{2001AJ....122.2833F,2003AJ....125.1649F,2004AJ....128..515F,2005astro.ph.12080F},
show a complete absorption trough blueward of the redshifted \Lya\
line from scattering by neutral hydrogen in the intergalactic medium
\citep{2001AJ....122.2850B}, the so-called Gunn-Peterson effect
\citep{1965ApJ...142.1633G}.  However, because of the extreme
sensitivity of the attenuation to the neutral fraction of \ion{H}{1},
it is only possible to deduce that $z \sim 6$ signals the end of the
epoch of reionisation (or even the end of the most recent epoch of
reionisation; e.g., \citealt{2003ApJ...586..693W}).

One way of robustly determining the actual epoch of reionisation is to
measure the luminosity function of the \Lya\ flux of high redshift
galaxies, which will be attenuated by the damping wing of the
Gunn-Peterson trough in a less severe manner than the continuum flux
blueward of \Lya.  Several searches for \Lya\ emitters (LAEs) at $z
\sim 6.5$ have been made, yielding a (currently published) total of 13
that have been spectroscopically confirmed
(\citealt{2002ApJ...568L..75H}; \citealt{2003PASJ...55L..17K} --- two;
\citealt{2004ApJ...611...59R}; \citealt{2004A&A...422L..13K};
\citealt{2005ApJ...619...12S}; \citealt{2005PASJ...57..165T} --- seven
new).  These sources show little evolution from analagous populations
at $z \sim 5.7$ \citep{2005ApJ...619...12S}, perhaps indicating that
we have not yet reached the epoch of reionisation.  Studies using
color break samples show a drop in the star formation rate by about a
factor of five at $z\sim 6$ from the peak at lower redshifts
\citep{2004MNRAS.355..374B,2005astro.ph..9641B}, while the \Lya\
surveys appear to show a flatter evolution \citep{hu-review}.

The color break searches are restricted to the small deep fields
observed with HST, while, the searches for LAEs are intensive,
requiring large investments of premier ground-based facilities, and
they select only the most luminous LAEs.  These surveys may therefore
be biased towards detecting the more vigorously star-forming galaxies
that produce large Str\"omgren spheres which allow the \Lya\ flux to
leak out.  This means that the measurement of the epoch of
reionisation will be biased towards higher redshifts.  To address this
bias, we require a means of homogeneously identifying more modest
star-forming galaxies at high redshift --- this is what gamma-ray
bursts (GRBs) may provide.

The luminosity function of the optical afterglows of GRBs extends as
bright as an absolute magnitude of $M_{R} \approx -31.5$ mag at 1~hour
after the GRB in the rest-frame, and probably even brighter at earlier
epochs.  Because of this extreme luminosity, they can be detected to
great distances and therefore provide an exciting way to find very
high-redshift galaxies beyond the current upper limits of $z \sim 7$
and to map the star formation history at these extreme redshifts in a
way which, if not itself unbiased, is at least independent of the
properties of the underlying galaxies \citep{2000ApJ...536....1L}.

Indeed, it was widely expected that the advent of the Swift satellite
would produce a large rate of return of very high redshift GRBs (e.g.,
\citealt{2002ApJ...575..111B}).  While the predictions appear, in
hindsight, to have been somewhat optimistic, the sensitive Swift
mission is detecting GRBs at a higher mean redshift of $z \sim 2$
\citep{2005astro.ph..5107B} than previous missions such as HETE-2,
BEPPO-SAX and IPN.  While this may make the afterglows more difficult
to identify, it also gave hope that SWIFT would find sources beyond
the most distant-known GRB ($z=4.5$; \citealt{2000A&A...364L..54A}).
The burst GRB~050904 discussed in this paper does just that, pushing
the redshift limit for GRBs beyond $z=6$.  Hopefully it is just the
first of many such detections stretching to still higher redshifts.

High redshift GRBs are easy to distinguish with coordinated optical
and near infrared (NIR) observations, since the Gunn-Peterson effect
drastically attenuates flux in the optical bands.  The mean
transmissions as a function of redshift based on quasar observations
to $z=6.4$ are tabulated in \citet{2004AJ....127.2598S}, and these can
be used to obtain the redshift of a GRB from its colors. However,
Swift's Ultra-Violet and Optical Telescope (UVOT) is limited to
observations at wavelengths bluer than about 6500\AA.  For this
reason, optical afterglows of GRBs at very high redshifts ($z>6$)
cannot be detected by the UVOT on Swift, and hence their properties
must be characterized using ground-based NIR observations.

GRB~050904 was triggered the Burst Alert Telescope (BAT) on Swift at
1:51 UTC on 2005 Sep.\ 4 and rapidly localized
\citep{2005GCN..3910....1C}.  \citet{2005GCN..3912....1F} undertook
optical observations in the optical $R$ and $i'$ bands starting about
3.5~hours after the GRB, but did not identify any afterglow candidate
to reasonable limiting magnitudes.  The subsequent identification of a
bright afterglow in the NIR $J$ band led to the interpretation that
this was a GRB at very high redshift \citep{2005astro.ph..9660H}.

Observations of the afterglow with the MAGNUM telescope began about 12
hours after the GRB (\S\ref{sec:magnum}).  These observations allow us
to generate a spectral energy distribution for the source at that time
which yields a strong lower limit on the redshift of $z=6.18$.  This
is consistent with the photometric redshifts reported by
\citet{2005astro.ph..9660H} and \citet{2005A&A...443L...1T}, and the
spectroscopic redshift of $z=6.295 \pm 0.002$ measured by
\citet{2005astro.ph.12052K}.  Furthermore, the limits on the break
strengths $R-J$ and $I-J$ are consistent with the object lying at the
spectrosopic redshift (\S\ref{sec:redshift}).  We describe these
observations in the present paper and briefly speculate on the
implications for the star formation history of the universe
(\S\ref{sec:discussion}).

\section{MAGNUM observations of GRB050904}
\label{sec:magnum}

MAGNUM (Multicolor Active Galactic NUclei Monitoring) is a 2~meter
telescope on Haleakala built by the Research Center for the Early
Universe (RESCEU) at the University of Tokyo and used to study AGN
variability \citep{yoshii}.  In order to optimise the efficiency of
the monitoring observations, the telescope is operated in a robotic
mode using queue scheduling.  GRB observations can be performed as
soon as a notification is received by inserting the target and
overriding the queue.

MAGNUM's principal instrumentation is the Multicolor Imaging
Photometer (MIP; \citealt{1998SPIE.3354..769K}), a dual-beam
optical/NIR camera which covers a 1.5~arcmin square field in the
$UBVRIYJHK_sKL'$ bands (though $U$ observations are difficult, and
$L'$ infeasible).  The instrument, mounted at the Cassegrain focus,
uses an internal beam splitter to send the short wavelength light to a
1024~pixel square thinned CCD (though the entire CCD is not
illuminated), and the long wavelengths to a 256~pixel square InSb
array.

Because the limited field of view of the MIP makes it impractical to
observe the 4~arcmin localisations from Swift's Burst Alert Telescope
(BAT), we target afterglows discovered by the X-Ray Telescope (XRT) or
ground-based follow-up observations in order to characterise their
spectral flux distribution, and so attempt to determine a photometric
redshift from the \Lya\ absorption at high redshift.  We use a
pre-planned sequence of four observations, each consisting of nine
individual minute-long exposures with a box dither pattern of
10~arcsec step: $RIRI$ in the optical and $HKJY$ in the NIR.  The
final images provide an accurate and nearly simultaneous spectral
energy distribution for the object.  The entire sequence takes about
84 minutes, with an additional 5 minutes before commencing the GRB
observations in order to correct the pointing and focus after the
slew.

\begin{figure*}
\centerline{\psfig{figure=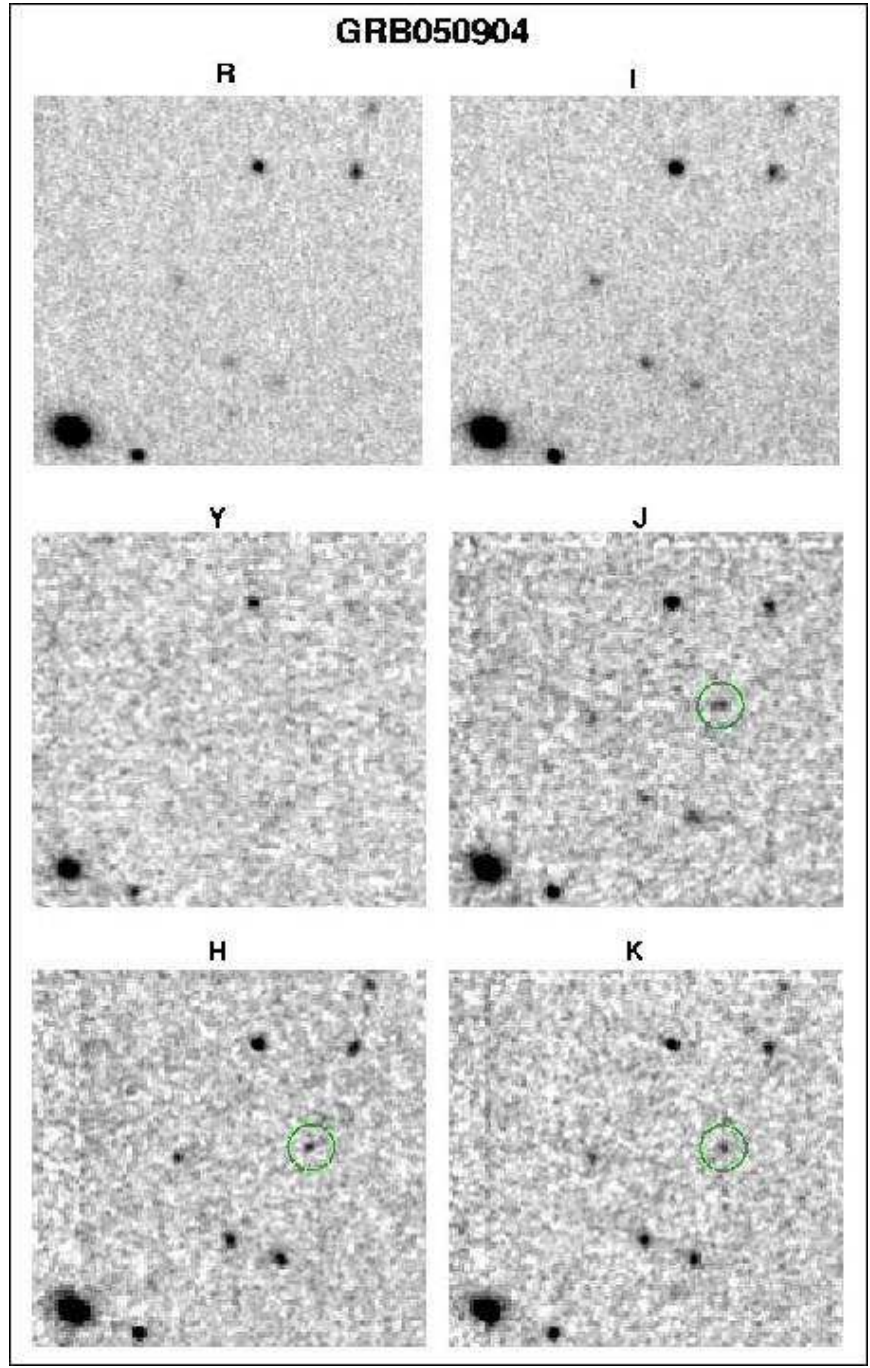,width=4.5in,angle=0}}
\caption{Images of GRB~050904 (circled object) in the $RIYJHK$ bands
respectively.  Each image is 1.2~arcmin on a side. The detector
response is poor at $Y$ and the failure to detect the afterglow at
this wavelength is not significant.}
\label{fig:images}
\end{figure*}

We observed GRB~050904 in this manner between 13:57 and 15:07 UTC on
2005 Sep.\ 4, or about 0.51 days after the GRB.  The final combined
images are shown in Figure~\ref{fig:images}.  Because the only 2MASS
source in the field is a galaxy, the images could not be immediately
photometrically calibrated.  We obtained observations on subsequent
nights to obtain $IJK$ calibrations of two stars in the field.  The
$H$ band was calibrated from the 2MASS galaxy by using a large
aperture to measure the entire flux of the galaxy, and applying a
measured aperture correction to obtain the flux of the afterglow.  The
$R$ band was calibrated using SDSS magnitudes for a star in the field,
and applying the appropriate transformation
\citep{2002AJ....123.2121S}.  The uncertainties in these two non-ideal
calibration methodologies do not strongly affect our results, since
the $R$ and $H$ measurements are less important than the $IJK$
measurements.  The fluxes in $\mu$Jy with 1~$\sigma$ errors are
summarized in Table~\ref{tab:fluxes}.  These fluxes are measured in
matched apertures corrected for the image quality, with radii ranging
from 1.4 to 1.7~arcsec. The $Y$ band observations are not particularly
sensitive, so are not included in our analysis.  Since no source is
apparent in the $RI$ bands, we measured the flux in the image at the
position of the afterglow.  Errors were taken as the r.m.s.\ from zero
of the distribution of flux in apertures randomly distributed over
empty background regions.

\begin{deluxetable}{lr}
\renewcommand\baselinestretch{1.0}
\tablecaption{Measured fluxes from MAGNUM observations in multiple bands.}
\tablehead{Band & Flux ($\mu$Jy)}
\startdata
$R$ & $-1.04 \pm 0.75$ \cr
$I$ & $0.4 \pm 1.2$ \cr
$J$ & $47.9 \pm 6.2$ \cr
$H$ & $30.0 \pm 8.9$ \cr
$K$ & $41 \pm 14$ \cr
\enddata
\tablecomments{The observations were made at a mean epoch of 2005
Sep. 4.60, and may be treated as simultaneous --- any correction to
the fluxes for the decay of the afterglow over the course of the
observations would be smaller than the measurement errors (0.15~mag,
using a temporal decay index of 1.2; \citealt{2005astro.ph..9660H}).
These measurements are not corrected for the relatively small
foreground Milky Way extinction.  The $R$ and $I$ band measurements
are consistent with no detections to $3\sigma$ upper limits on the
fluxes of 2.3 and 3.6~$\mu$Jy, respectively.}
\label{tab:fluxes}
\end{deluxetable}

The difference between the $J$-band magnitude reported here and that
obtained by UKIRT at a slightly earlier epoch, combined with the
unusual behaviour in the $Z$ band at about this epoch reported by
\citet{2005astro.ph..9660H} likely indicates that the source was
somewhat variable during the period of these observations.  Such short
timescale variability has been detected in other GRB optical
afterglows, with GRB~030329 a notable example
\citep{2004ApJ...606..381L}.  This serves to demonstrate the need for
simultaneous or near-simultaneous multicolor imaging, such as MAGNUM
provides.

\section{Redshift limits and the break strength}
\label{sec:redshift}

\begin{figure*}
\centerline{\psfig{figure=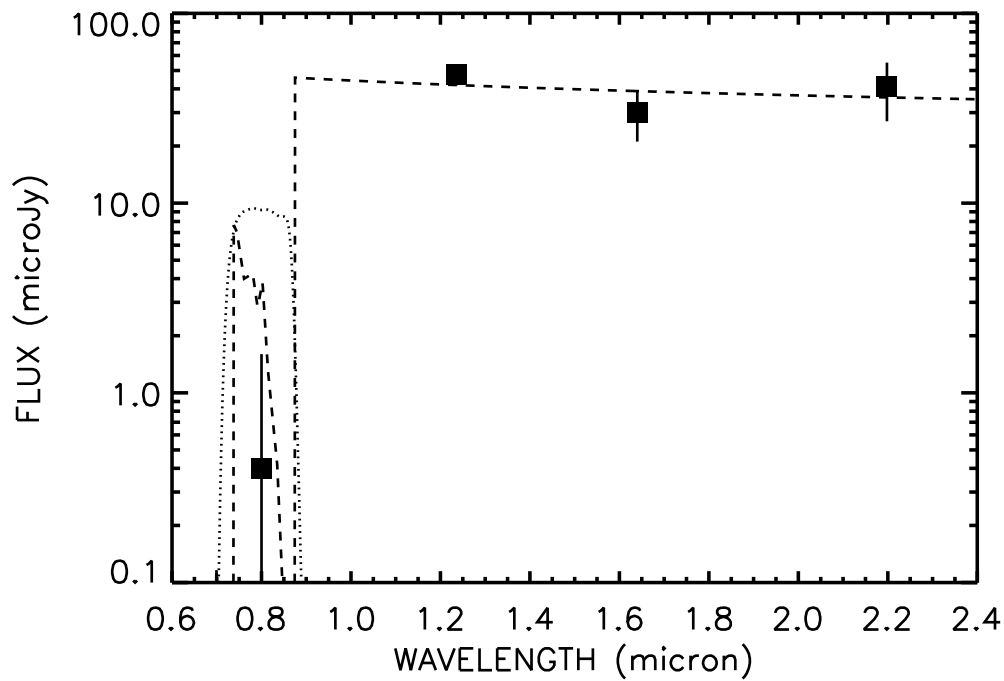,width=3.5in,angle=90}}
\caption{The spectral flux distribution of GRB~050904 from the MAGNUM
observations are shown by the solid squares with 1~$\sigma$ error
bars. The dotted line shows the position of the $I$ filter. The dashed
line shows the expected SFD of the object at a redshift of $z=6.18$
(our 2~$\sigma$ lower limit on the redshift) based on a power law fit
to the longer ($JHK$) wavelengths, modulated by the measured
transmission of the intergalactic medium below the redshifted \Lya\
position computed using the measurements of
\citet{2004AJ....127.2598S}.}
\label{fig:sfd}
\end{figure*}

We show the measured spectral flux distribution (SFD) of the source in
Figure~\ref{fig:sfd}.  There is a substantial break at the $I$ band
which places a tight lower limit on the redshift of the
source. Because the $I$ band data is consistent with a null detection
we cannot place any useful upper limit (better than $z \ale 8$) on the
redshift of the source.

In order to obtain the redshift estimate, we fitted a power-law
spectrum to the $JHK$ data, obtaining a spectral slope of $\beta =
0.3\pm0.6$, where $f_\nu \propto \nu^{-\beta}$; though the error bar
is large, such a shallow spectral slope is likely a product of
intrinsic variability in the source during the $J$-band observation,
as discussed earlier.  We then modulated this spectrum with the \Lya\
and Ly$\beta$ transmissions of the intergalactic medium measured by
\citet{2004AJ....127.2598S} in high redshift quasars.  The resulting
spectrum is shown for $z=6.18$ in Figure~\ref{fig:sfd}, where we also
show the positions of the $I$ filter.  In order to reduce the $I$ band
flux to observed value we require $z>6.18$ at the 2~$\sigma$ level and
$z>6.37$ at the 1~$\sigma$ level.  The results are extremely sensitive
to the adopted redshift.  They are consistent with the redshift range
of $z = 6.30 \pm 0.07$ reported by \citet{2005A&A...443L...1T} based
on similar observations with the VLT and also with the spectroscopic
redshift of $z=6.295 \pm 0.002$ found by \citet{2005astro.ph.12052K}.

\begin{figure*}
\centerline{\psfig{figure=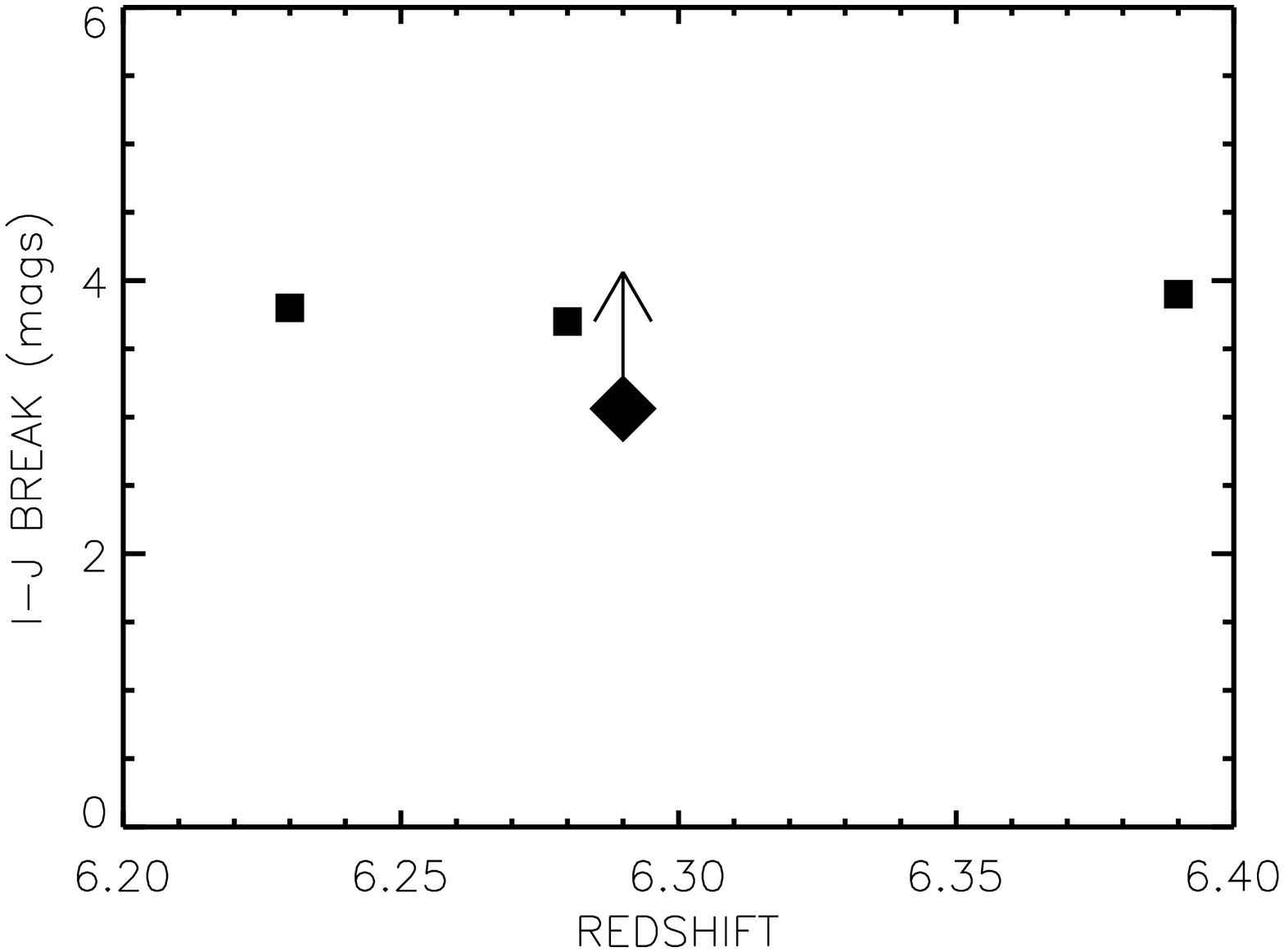,width=3in,angle=90}}
\centerline{\psfig{figure=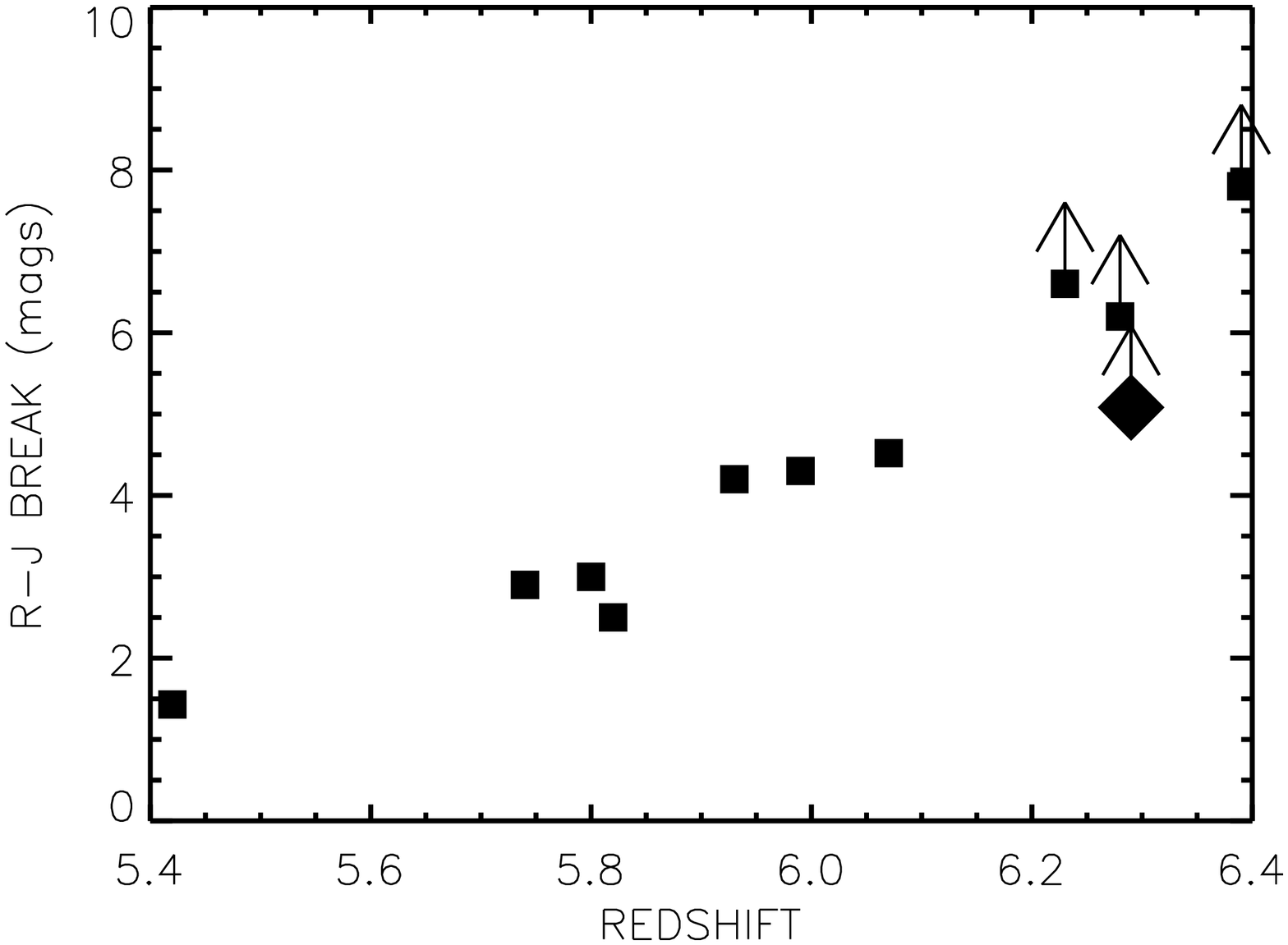,width=3in,angle=90}}
\caption{2~$\sigma$ lower limits on the $I-J$ and $R-J$ break measured
in the GRB (solid diamonds) are compared with directly measured values
in high redshift quasars (solid squares).}
\label{fig:breaks}
\end{figure*}

In Figure~\ref{fig:breaks} we show the break strength between $R$ and
$J$, and between $I$ and $J$, directly compared with the measured
values in individual quasars at these redshifts. The quasar values are
measured by comparing a power law fit to the continuum in line free
regions of the quasar to the directly measured flux in the $I$ band
\citep{2004AJ....127.2598S}. The $R$ band provides a weaker constraint
than the $I$ band but would still place the GRB at $z>6.1$.

%

\section{Discussion}
\label{sec:discussion}

Observations of high redshift GRBs can be used to infer the
cosmological star formation history either through observations of the
host galaxy or by translating the GRB rate to a star formation rate
\citep{1997ApJ...486L..71T}.  The most direct method is to assume that
the rate of GRBs as a function of redshift is proportional to the rate
of formation of high-mass stars and hence (assuming that the mass
function is invariant) to the total star formation rate.  In order to
make such an interpretation, we need to calibrate the GRB rate versus
the star formation history at lower redshifts, and it will always be
subject to the assumption that the fraction of massive stars forming
GRBs and the initial mass function of the stars remain invariant at
the higher redshifts (e.g., \citealt{2001ApJ...548..522P};
\citealt{2000ApJ...536....1L}; \citealt{2002ApJ...575..111B};
\citealt{2005astro.ph..5496N}).


Prior to GRB~050904 twenty spectroscopic redshifts had been measured
for Swift GRBs (excluding short/hard GRBs), of which fourteen lie
beyond $z=1$. Thirteen of these fourteen would still have been
detected above the Swift BAT threshold of 0.2~ph/cm$^2$/s if they had
lain at $z=6.29$.  This indicates that the selection effects, at least
in the gamma-rays, are not strong.  Here we make the simple assumption
that the thirteen GRBs summarized in Table~2 which would have fluxes
about 0.2 ph/cm$^2$/s at $z=6.29$ represent the low redshift
counterparts of GRB~050904 detected over the same period of Swift
observations, which indicates that the efficiency of detecting GRBs at
$z=6.29$ is near unity, compared to GRBs at lower redshifts. For each
GRB we give the redshift, the observed peak flux and the value of the
peak flux if the source had been at $z=6.29$.

\begin{deluxetable}{crrrc}
\renewcommand\baselinestretch{1.0}
\tablecaption{Lower redshift Swift GRBs counterparts to GRB050904}
\tablehead{GRB & Redshift & Peak flux & $f_{z=6.29}$ & References}
\startdata
051111  & 1.549 & 2.50 & 0.25 & (1), (2) \\
051109A & 2.346 & 3.70 & 0.77 & (3), (4) \\
050922C & 2.198 & 7.36 & 1.37 & (5), (6) \\
050908  & 3.350 & 0.70 & 0.27 & (7), (8) \\
050820A & 2.612 & 2.50 & 0.63 & (9), (10) \\
050802  & 1.710 & 2.65 & 0.32 & (11), (12) \\
050730  & 3.969 & 0.57 & 0.28 & (13), (14) \\
050603  & 2.821 & 27.6 & 7.89 & (15), (16) \\
050505  & 4.270 & 1.81 & 1.01 & (17), (18) \\
050401  & 2.900 & 12.6 & 3.77 & (19), (20) \\
050319  & 3.240 & 1.45 & 0.52 & (21), (22) \\
050318  & 1.440 & 3.20 & 0.28 & (23), (24) \\
050315  & 1.949 & 1.98 & 0.30 & (25), (26) \\
050126  & 1.290 & 0.70 & 0.05 & (27), (28) \\
\enddata

\tablecomments{The sample is limited to Swift GRBs at $z>1$.  The peak
fluxes are in ph/cm$^2$/s, measured in the 15--150~keV band; they are
taken from the Swift
archive\footnote{http://swift.gsfc.nasa.gov/docs/swift/archive/grb\_table.html}.
All but GRB~050126 would have been detected by Swift if placed at
$z=6.29$.  References: (1) \citet{2005GCN..4255....1H}; (2)
\citet{2005GCN..4260....1K}; (3) \citet{2005GCN..4221....1Q}; (4)
\citet{2005GCN..4217....1F}; (5) \citet{2005GCN..4029....1J}; (6)
\citet{2005GCN..4020....1K}; (7) \citet{2005GCN..3948....1F}; (8)
\citet{2005GCN..3951....1S}; (9) \citet{2005GCN..3833....1P}; (10)
\citet{2005GCN..3835....1C}; (11) \citet{2005GCN..3749....1F}; (12)
\citet{2005GCN..3737....1P}; (13) \citet{2005GCN..3709....1C}; (14)
\citet{2005GCN..3715....1M}; (15) \citet{2005GCN..3520....1B}; (16)
\citet{2005GCN..3512....1F}; (17) \citet{2005GCN..3368....1B}; (18)
\citet{2005GCN..3364....1H}; (19) \citet{2005GCN..3176....1F}; (20)
\citet{2005GCN..3173....1S}; (21) \citet{2005GCN..3136....1F}; (22)
\citet{2005GCN..3134....1K}; (23) \citet{2005GCN..3122....1B}; (24)
\citet{2005GCN..3119....1K}; (25) \citet{2005GCN..3101....1K}; (26)
\citet{2005GCN..3105....1K}; (27) \citet{2005GCN..3088....1B}; (28)
\citet{2005GCN..2987....1S}.  }
\label{tab:bursts}
\end{deluxetable}

We point out that this is, of course, a simplified analysis and that
there are more selection effects than just the detection of the GRB
itself.  In particular, the success rate of detecting the optical/NIR
afterglow will be a function of redshift, as will the fraction of
afterglows for which it is possible to measure the redshift, either
from absorption or emission lines.  This is a complicated endeavour
requiring detailed Monte Carlo simulations of GRB afterglow searches
to determine which GRBs with and without redshifts might plausibly
have been identified if followed up in the same way as GRB~050904, and
is beyond this simplified analysis.  However, such an analysis is not
justified at the present time, given the significant small number
uncertainties in the data and the desire for a larger sample of Swift
events to use in bootstrapping the completeness estimate.
Nevertheless, our measurement will provide, at the least, a lower
limit to the star formation density, since we know that we are missing
some GRBs.

Because of these complicated selection effects, which have discouraged
all but the most bold (e.g., \citealt{2000MNRAS.312L..35B}) from using
the actual redshifts measured from the optical afterglow or host
galaxy in determining the GRB rate as a function of redshift, most
other attempts have relied upon assuming that GRBs are standard
candles (e.g., \citealt{1999ApJ...511...41T}), or using empirical
``pseudo-redshift'' indicators from the GRB itself (e.g.,
\citealt{2004ApJ...611.1033F,2002ApJ...574..554L}) such as the
luminosity-variability relation \citep{2001ApJ...552...57R}.
These methods buy a large sample size and better understood selection
effects at the cost of uncertain (by at least a factor of two!)
redshifts.  This results in a need to deconvolve the resulting rate
distribution using the redshift errors, making it difficult to
determine the true behaviour of the GRB rate at high redshift.

The relation between the star formation history and the GRB rate has
been derived by a number of authors
(e.g. \citealt{2001ApJ...548..522P}; \citealt{2000ApJ...536....1L};
\citealt{2002ApJ...575..111B}; \citealt{2005astro.ph..5496N}).  At
$z>1$ where the effects of the cosmological constant are negligible,
the number of gamma ray bursts per unit redshift, $dN/dz$, in a given
observing time interval is simply related to the star formation rate
per unit comoving volume, $\psi$, by the function:
\begin{equation}
dN/dz = A \, \psi \, [(1+z)^{-2.5} - 2(1+z)^{-3} + (1+z)^{-3.5}]
\end{equation}
where the normalizing factor, $A$, is assumed to be independent of
$z$.

\begin{figure*}
\centerline{\psfig{figure=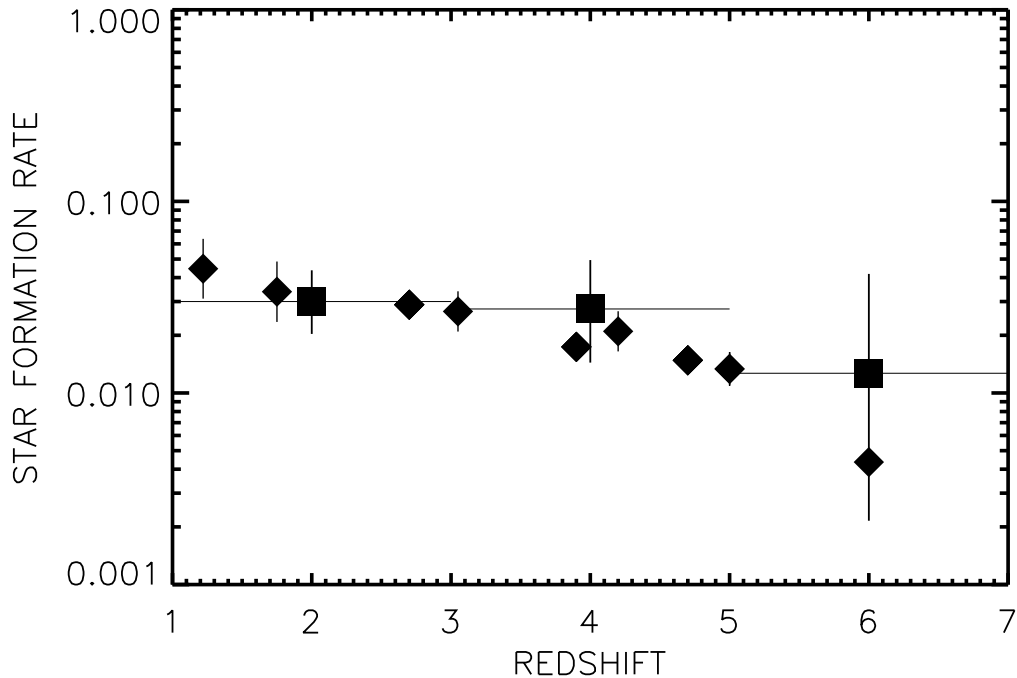,width=3.5in,angle=90}}
\caption{The star formation rate as a function of redshift in units of
solar masses per Mpc$^{3}$, taken from the compilation of
\citet{2004MNRAS.355..374B} is shown by diamonds.  The solid boxes
with 1~$\sigma$ error bars show the corresponding determinations from
the Swift GRBs for a normalizing constant of 0.0033~in the same
units.}
\label{fig:sfr}
\end{figure*}

In Figure~\ref{fig:sfr} we compare the shape of the star formation
rate determined from the Swift GRBs with color-selected galaxy
determinations of the star formation rate over the same redshift range
taken from the paper of \citet{2004MNRAS.355..374B}.  We have set
$A=0.005$~M$_\odot$~Mpc$^{-3}$ to match the observations at $z=3$.
Within the wide uncertainties left by the small number of statistics,
the current values cannot differentiate between the slow decline seen
in the color selected galaxies and a flat star formation rate with
redshift.  However, it is clear that as the sample size increases we
should be able to make a valuable comparison.  The GRB determinations
are more powerful in some ways, since they relate to all star
formation including those in lower luminosity galaxies than can be
directly detected at the present time.  This holds promise that future
identifications of $z>6$ GRBs will enable a complete measurement of
the star formation rate density at very high redshift.

\section{Summary}
\label{sec:summary}

In the present paper we have described the observations of GRB~050904
with the MAGNUM telescope. These observations place a strong lower
limit on the redshift of $z=6.18$ consistent with the spectroscopic
redshift of $z=6.29$ measured by \citet{2005astro.ph.12052K}.

The most immediate result is that GRBs exist at $z > 6$, and that they
can be identified using a simple set of near-simultaneous optical and
NIR observations.  This presents the prospect of using the afterglows
of high-redshift GRBs not only as lighthouses to illuminate the
high-redshift Universe (as is done for quasars today), but also as
signposts to alert observers to the presence of the host galaxy,
allowing deep follow-up observations to measure the \Lya\ flux.  With
the discovery of more high-redshift GRBs, it should be possible to
form a useful sample for nailing down the epoch of reionisation.

We also gave a simple discussion of the star formation rate history
from $z=1-7$ based on the current Swift GRB observations showing that
within the still broad uncertainties, the observations point to a flat
or a slowly declining star formation rate consistent with color
selected galaxy observations.

GRB~050904 is an exciting precursor to further high redshift GRBs
which should allow us to refine the star formation analysis and to
study the properties of the intergalactic medium at these redshifts
through color break and spectroscopic techniques.

\acknowledgments

We thank Elizabeth Stanway for providing the data for
Figure~\ref{fig:sfr} in tabular form.  This work was supported by a
Swift Guest Investigator grant (NNG05GF40G) and a Grant in Aid of
Center of Exellence Research (07CE2002) of the Ministry of Education,
Science, Culture and Sports of Japan.  We thank the anonymous referee
for helpful suggestions that improved this paper.


\end{document}